\documentclass[aps,floats,twocolumn,prb,showpacs]{revtex4-1} 
\usepackage{graphicx}
\usepackage{amsmath,amssymb}
\usepackage{subfigure}
\usepackage[version=3]{mhchem}
\usepackage{xspace}
\usepackage[utf8]{inputenc}
\usepackage{tabularx}
\usepackage[colorlinks, allcolors=black]{hyperref}

\newcommand\svo{\ce{SrVO3}\xspace}
\newcommand\boo{\ce{BaOsO3}\xspace}
\newcommand\ttg{\ensuremath{t_{2g}}\xspace}
\newcommand\eg {\ensuremath{e_{g}}\xspace}
\newcommand\RR {\ensuremath{\mathbf R}\xspace}
\newcommand\kk {\ensuremath{\mathbf k}\xspace}
\newcommand\oo {o'}
\newcommand\dd {d'}

\begin{document}

\title{Cubic interaction parameters for $t_{2g}$ Wannier orbitals}

\author{T. Ribic}
\author{E. Assmann}
\author{A. T\'{o}th}
\author{K. Held}
\affiliation{Institute of Solid State Physics, Vienna University of
  Technology, A-1040 Vienna, Austria}

\date{\today}

\begin{abstract}
  Many-body calculations for multi-orbital systems at present
  typically employ Slater or Kanamori interactions which implicitly
  assume a full rotational invariance of the orbitals, whereas the
  real crystal has a lower symmetry.  In cubic symmetry, the
  low-energy $t_{2g}$ orbitals have an on-site Kanamori interaction,
  albeit without the constraint $U=U'+2J$ implied by spherical
  symmetry ($U$: intra-orbital interaction, $U'$: inter-orbital
  interaction, $J$: Hund's exchange).  Using maximally localized
  Wannier functions we show that deviations from the standard,
  spherically symmetric interactions are indeed significant for $5d$
  orbitals ($\sim 25\%$ for \boo; $\sim 12\%$ if screening is
  included), but less important for $3d$ orbitals ($\sim 6\%$ for
  \svo; $\sim 1\%$ if screened).
\end{abstract}

\maketitle
\section{Introduction}

Strongly correlated electron systems show a rich variety of
unconventional phenomena such as high temperature superconductivity
\cite{highTc} and quantum criticality \cite{qucrit} --- and their
theoretical description and understanding constitutes a particular
challenge.  The origin of these correlations is the strong Coulomb
interaction, as particularly found in materials with partially filled
$d$- or $f$-bands, such as transition metals, their oxides, rare earth
and lanthanide compounds.

The Coulomb interaction between two electrons, which scatter from
orbitals $\alpha$, $\beta$ to $\alpha'$, $\beta'$ in the course of the
interaction, is simply given by
\begin{multline}
U_{\alpha' \beta' \beta \alpha} =\\ \int \!{\rm d}^3r {\rm d}^3r'
\psi_{\alpha'}^*({\mathbf r}) \psi_{\beta'}^*({\mathbf r'}) 
V(|{\mathbf r'}\!-\!{\mathbf r}|)
\psi_{\beta}({\mathbf r'})  \psi_{\alpha}({\mathbf r}) \;.
\label{Eq:U}
\end{multline}
Here, $V(|\mathbf r'\!- \mathbf r|) = e^2 / \big(4\pi \epsilon_0
|\mathbf r'\!- \mathbf r|\big)$ is the Coulomb interaction with
electron charge $e$ and vacuum permittivity $\epsilon_0$;
$\psi_{\alpha}({\mathbf r})$ is the electron wave function for orbital
$\alpha$; no screening by further electrons has been included in this
bare interaction $U_{\alpha' \beta' \beta \alpha}$.  We do not
consider relativistic corrections such as the spin-orbit coupling here
so that the one-electron eigenstates simply need to be multiplied with
a spinor and the integrals $U_{\alpha' \beta' \beta \alpha}$ are
independent of spin; the $\alpha'$ and $\alpha$ one-electron
eigenstates (as well as $\beta'$ and $\beta$) need to have the same
spin though.

For practical calculations, it is essential to reduce the number of
interaction parameters. Often, e.g.\ in DFT+$U$ (density-functional
theory augmented by a Hubbard-$U$ interaction in a static mean-field
approximation) \cite{DFTU} and DFT+DMFT (dynamical mean-field theory),
\cite{DFTDMFT} one considers only the \emph{local} interaction. That
is, all orbitals $\alpha, \beta$ in Eq.\ (\ref{Eq:U}) are on the same
site; they might correspond to Wannier orbitals \cite{WannierOrig}
localized around the same lattice site. This is justified not only
because this on-site interaction is by far the largest interaction
parameter, but also since non-local interactions between orbitals on
different sites can be treated in simple (Hartree) mean field theory
in the limit of a large number of neighbors. \cite{MH} Certainly there
are situations where such non-local interactions can be of importance,
particularly in one- and two-dimensions, or also between transition
metal $d$ and oxygen $p$ orbitals.  \cite{Nico}

A further reduction of parameters can be achieved using the so-called Slater
integrals \cite{Slater}
\begin{equation}
F_l=\int\!  {\rm d}r {\rm d}r' {R(r)}^2 {R(r')}^2 \dfrac{ {\rm
    min}(r,r')^l }{  {\rm max}(r,r')^{l+1} } \ r^2  \ {r'}^2\,.
\label{Eq:Slater}
\end{equation}
Here, the underlying assumption is spherical symmetry, which allows
for an analytical angular integration so that eventually only the
integrals Eq.\ (\ref{Eq:Slater}) over the radial part $R(r)$ of the
wave functions remain, see Appendix.  These Slater integrals, the
simpler Kanamori \cite{Kanamori} interaction, and or even just a
single $U$-parameter are commonly used in DFT+$U$,\cite{DFTU , LDAUU}
DFT+DMFT,\cite{DFTDMFT,LDADMFTU,LDADMFTU2} or full-multiplet
configuration-interaction calculations.  \cite{fullmultiplet,
  Haverkort12, notecRPA} However, a crystal lattice is not spherically
symmetric.  It has a lower, e.g.\ cubic, symmetry.

The aim of our paper is hence to analyze the nature and magnitude of
the deviations from spherical interaction parameters. To this end, we
study the specific and arguably most relevant case of transition metal
oxides with a cubic perovskite (\ce{$AB$O3}) structure.  In Section
\ref{Sec:analytic}, we study analytically the structure of the Coulomb
matrix elements for a $B$O$_6$ octahedron.  For the low energy \ttg
orbitals, the cubic Coulomb interaction requires three parameters
instead of the two parameters for spherical symmetry. We explicitly
derive the most relevant integrals that deviate from the Slater
integrals (\ref{Eq:Slater}).

In Section \ref{Sec:numeric}, we calculate the quantitative deviations
from spherical symmetry by means of maximally localized Wannier
orbitals.  While the bare interaction in $3d^1$ \svo is still
described reasonably well by spherically symmetric interaction
parameters, the stronger $p$-$d$ hybridization in $5d^4$ \boo results in
larger deviations ($\sim 25\%$). In a Wannier basis which includes
both the transition metal \ttg and the oxygen $p$ orbitals, working
with spherically symmetric interactions is justified.  Even for \boo
deviations between cubic and spherical symmetric interactions are only
$3\%$ in this case.

The effect of screening within the Thomas-Fermi approximation is
considered in Section \ref{Sec:screening}.  For short screening
lengths, deviations from spherical symmetry are even larger than in
the unscreened case; for realistic screening lengths, deviations are
reduced but still significant for \boo in a 3-orbital Wannier basis
($\sim 12\%$).

\section{Cubic interaction parameters}
\label{Sec:analytic}
\begin{figure}
  \raisebox{-.5\height}{\includegraphics[width=.5\linewidth]{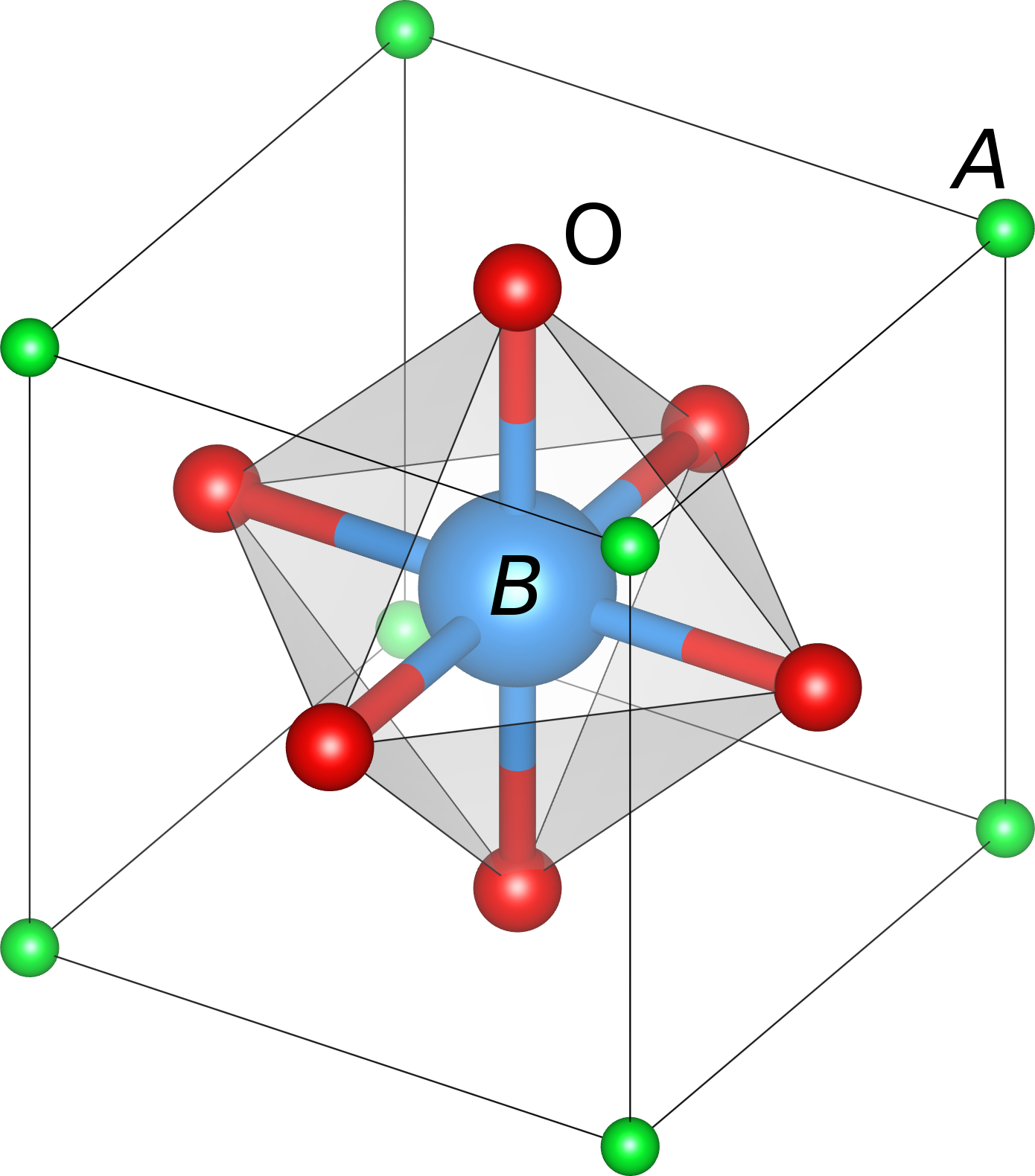}}
  \hfill
  \raisebox{-.5\height}{\includegraphics[width=.4\linewidth]{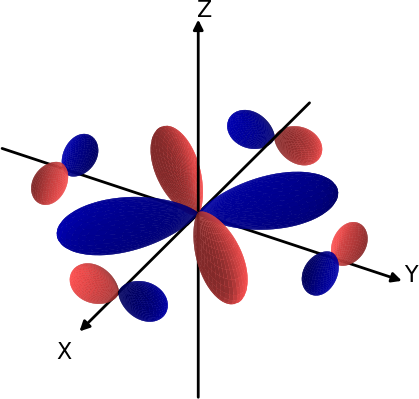}}
  \caption{(Color online.)  Left: In the perovskite (\ce{$AB$O3})
    structure, the oxygen octahedron around the central transition
    metal ion breaks spherical symmetry down to cubic.  The \ce{$B$}
    ion (large sphere) occupies the center of a cube with \ce{$A$} at
    the corners and \ce{O} at the face centers.  Right: Schematic
    representation of the low energy $\dd_{xy}$ orbital of
    Eq.~(\ref{Eq:dprime}) [light / dark shading indicates opposite
    signs of the wave function].}
  \label{Fig:octahedron}
\end{figure}

We consider the typical situation for transition metal oxides with an
octahedron of oxygens surrounding each transition metal atom as shown
in Fig.~\ref{Fig:octahedron}.  While an isolated transition metal atom
would be spherically symmetric and the parameterization in terms of
Slater integrals exact, the oxygen octahedron reduces the symmetry to
cubic point group symmetry \cite{Koster} around the transition metal
atom.  Therefore, the fivefold degeneracy of the atomic $d$ level is
partially lifted, leaving a threefold degenerate \ttg and a twofold
degenerate \eg level in the cubic environment.  In the cases we
consider, the octahedron vertices are occupied by negatively charged
\ce{O^{2-}} ions.  In this case, the \eg states, which have a lot of
weight along the \ce{$B$}--\ce{O} lines, are higher in energy than the
\ttg states, whose weight resides predominantly in the space between
the \ce{O} ions, see Fig.~\ref{Fig:octahedron} (right).

The effective \ttg orbitals are a combination of predominantly
transition metal $d$ orbitals admixed with oxygen $p$ orbitals. For
many transition metal oxides, these \ttg orbitals constitute the
low-energy degrees of freedom for excitations around the Fermi
energy.\cite{MIT} For an analytical description we consider an atomic
transition metal \ttg orbital, denoted as $d_{\alpha}$ with
$\alpha\in\{xy,yz,xz\}$ in the following. This $d_{\alpha}$ orbital
mixes with a linear combination of oxygen $p$ orbitals of the same
symmetry, see, e.g., Ref.~\onlinecite{Haverkort12}.  It is convenient
to define this linear combination as $o_{\alpha}$: e.g.,
\begin{equation}
o_{xy} = (p^{+y}_x + p^{+x}_y - p^{-y}_x - p^{-x}_y)/2 \; ,
\label{Eq:Oxy}
\end{equation}
where $p^{+y}_x$ is the $p_x$ orbital centered around the oxygen atom
in the the positive $y$ direction, see Fig.\ \ref{Fig:octahedron}.
The orbitals $o_{xz}$ and $o_{yz}$ follow from Eq.~(\ref{Eq:Oxy}) by
cubic symmetry, i.e., $x \leftrightarrow z$ and $y \leftrightarrow z$,
respectively.

Symmetry ensures that the orbital $o_{\alpha}$ is orthogonal to
$d_{\alpha'}$, except when $\alpha=\alpha'$.  Thus to orthogonalize
the set of orbitals $\{o_{\alpha},d_{\alpha'}\}$, one has only to
orthonormalize $o_{\alpha}$ with respect to its associated
$d_{\alpha}$.  The orthonormalized orbitals are
\begin{equation}
\oo_{\alpha=ij} = \dfrac{o_{ij} -  d_{ij}\,  \langle d_{ij} | o_{ij}
  \rangle  }{ \sqrt{ 1 - \langle d_{ij} | o_{ij} \rangle}}
\,,
\end{equation}

The mixing of transition metal $d$ orbitals and oxygen $p$ orbitals
stems from hybridization; by symmetry, there is 
a hybridization only between $d_{\alpha}$ and $\oo_{\alpha}$ with the same
$\alpha$. Hence we obtain the tight-binding Hamiltonian
\begin{equation*}
  \begin{pmatrix}
    E_d & t \\
    t & E_p \\
  \end{pmatrix}
\end{equation*}
where $E_d$ and $E_p$ are the $d$ and $p$ (more precisely the
orthogonalized $\oo$) energy level; $E_d-E_p$ is the charge transfer
energy.  The predominantly $d$ eigenfunctions of this tight-binding
Hamiltonian, $\dd_{ij}$, are the effective low energy \ttg orbitals
\begin{equation}
\dd_{ij} = a  d_{ij} + b \oo_{ij}
\label{Eq:dprime}
\end{equation}
with $\eta = (E_d - E_p)/2t$, $a = \bigl[2 ( \eta^2 - \eta
  \sqrt{\eta^2 + 1} +1 ) \bigr]^{-1/2}$, and $b = \bigl[2 ( \eta^2 + \eta
    \sqrt{\eta^2 + 1} +1 ) \bigr]^{-1/2}$. 

After defining the low energy \ttg orbitals, we need to calculate the
Coulomb interaction between these one-particle eigenstates, i.e.,
\begin{equation}
U_{\alpha' \beta' \beta \alpha} =
\langle \dd_{\alpha'=ij} | \langle \dd_{\beta'=kl} |V | \dd_{\beta=mn} \rangle | \dd_{\alpha=op} \rangle \; .
\end{equation}
This is the relevant site-local Coulomb interaction for the low energy
degrees of freedom.  Note that in this context, $U_{\alpha' \beta'
  \beta \alpha}$ is defined as matrix element between direct products
of single-particle states denoted as $ | \dd_{\beta=mn} \rangle |
\dd_{\alpha=op} \rangle$, not between antisymmetrized Fock-states.

Since often $b\ll 1$ in transition metal oxides, we consider in the
following only the leading terms in the limit of large distance
between transition metal and oxygen site. In this limit, the direct
overlap $\langle d_{ij}| o_{ij}\rangle$, $b$ (which is the overlap
with respect to the one-particle Hamiltonian, $b \sim t$), and Coulomb
integrals between orbitals on different sites are small.  In the
following we hence restrict ourselves to all terms of up to second
order in (any) of the above off-site overlaps, and obtain the
following three contributions:

Directly from the $a  d_{ij}$ terms in Eq.~(\ref{Eq:dprime}) and
from the orthogonalization of the $\oo_{ij}$ we get a contribution
\begin{equation}
 \big( a^4 - 4 a^3 b \dfrac{\langle d_{uv}| o_{uv} \rangle}{N}  \big) \langle {d}_{ij} | \langle {d}_{kl} | V | {d}_{mn} \rangle | {d}_{op} \rangle
\label{Eq:dddd}
\end{equation}
This term is centered around the transition metal ion and can be
expressed in terms of the Slater integrals $F_l$ for the $d_{ij}$
orbitals.  Hence, this term can still be parameterized with Kanamori
interaction parameters.

From two $b \, \oo_{ij}$'s  in Eq.~(\ref{Eq:dprime}) we get a contribution
\begin{equation}
  \big(2 a^2 b^2 \dfrac{1}{N^2}  \big) \langle {d}_{ij} | \langle {o}_{kl} | V | {o}_{mn} \rangle | {d}_{op} \rangle \;.
\label{Eq:ppdd}
\end{equation}
Note, ${o}_{kl}$ and ${o}_{mn}$ have a contribution from the same
oxygen site, so that the $r$ and $r'$ integrals both include on-site
overlaps. Since for large oxygen-transition metal distances the
inter-site overlap decays exponentially while the Coulomb interaction
decays like $1/r$, we keep the term Eq.~(\ref{Eq:ppdd}).

Finally, there is a contribution involving only one $b \, \oo_{ij}$ in
Eq.~(\ref{Eq:dprime}) and a Coulomb integral overlap between
transition-metal and oxygen site:
\begin{equation}
  \big(2 a^3 b  \dfrac{1}{N}  \big)  \langle {d}_{ij} | \langle {d}_{kl} | V | {o}_{mn} \rangle | {d}_{op} \rangle \;.
\label{Eq:pddd}
\end{equation}
All other terms are of higher order in $b$ or the off-center overlap
integrals.

Eqs.~(\ref{Eq:ppdd}) and (\ref{Eq:pddd}) involve Coulomb integrals with
two distinct sites, oxygen and transition metal. Hence, 
they cannot be expressed in terms of Slater integrals any longer.
One can also envisage that from the orbital in Fig.~\ref{Fig:octahedron}
(right). While the spherical rotations around the $x$ or $y$ axis of
 the central  $d_{xy}$ part of the 
$\dd_{xy}$ orbital in  Fig.~\ref{Fig:octahedron}
(right) map the $d_{xy}$ orbital onto a linear combination of the three $d_{\alpha}$
orbitals, this is not possible any longer with the oxygen admixture
$\oo_{xy}$ in $\dd_{xy}$ except for 90 degrees rotations. Non-cubic rotations will put the rotated orbitals into positions where there is actually no oxygen site.

Employing the cubic symmetry, we can further reduce the number of
integrals needed in Eqs.~(\ref{Eq:dddd}), (\ref{Eq:ppdd}) and
(\ref{Eq:pddd}); or (\ref{Eq:U}) cf.~\onlinecite{Georges13}.  Any
integral involving an orbital index $\alpha=ij$ once or thrice is odd
in one cubic direction and hence vanishes.  This leaves us with
integrals where all orbitals are the same, i.e., the intra-orbital
Hubbard interaction $U=U_{\alpha\alpha \alpha\alpha}$ and integrals
where we have two distinct orbitals $\alpha\neq \beta$ twice. For the
latter we have the three possibilities: the inter-orbital interaction
$U'=U_{\alpha\beta \beta\alpha}$, the Hund's exchange
$J=U_{\alpha\beta \alpha\beta}$, the pair hopping term and $U_{\alpha
  \alpha \beta\beta}$ which for real-valued wave functions has the
same amplitude as $J$.  These symmetry considerations actually hold in
general, but without spherical symmetry $U\neq U'+2J$ because of the
terms Eqs.~(\ref{Eq:ppdd}) and (\ref{Eq:pddd}).  For spherical
symmetry, the connection to the Slater integrals is as follows,
cf.~\onlinecite{Georges13}: $U=F_0+\frac{4}{49}(F_2+F_4)$,
$U'=F_0-\frac{2}{49} F_2 - \frac{4}{441}F_4$,
$J=\frac{3}{49}F_2+\frac{20}{441}F_4$, so that $U=U'+2J$
holds. \cite{rotnote} If we have instead only cubic symmetry, we can
still parameterize the interaction in terms of $U$, $U'$, and $J$, but
now with $U \neq U'+2J$ and no expression in terms of Slater
integrals.

In second quantization, this Kanamori Hamiltonian,\cite{Kanamori}
which is obtained from Eq.(\ref{Eq:U}) by including all valid spin
combinations in Eq.(\ref{Eq:U}), reads:
\begin{multline} 
{H}_U =
\frac12 \sum_{\substack{\alpha, \beta\\ \alpha^\prime, \beta^\prime}}
U_{\alpha^\prime \beta^\prime \beta
\alpha}\,\,{\sum_{\sigma, \sigma^\prime}}
\,\,c^{\dagger}_{\alpha^\prime \sigma}
\,\,c^{\dagger}_{\beta^\prime \sigma^\prime}\,\,c^{}_{\beta
  \sigma^\prime}\,\,c^{}_{\alpha \sigma} 
\\
= U \sum_{\alpha} n_{\alpha,\uparrow} n_{\alpha,\downarrow} +
\sum_{\substack{\alpha>\beta\\\sigma,\sigma'}} \Big[
(U'-\delta_{\sigma \sigma'}J) n_{\alpha,\sigma} n_{\beta,\sigma'}
\Big] 
\\
-\sum_{\alpha \neq \beta}
J(c^\dagger_{\alpha,\downarrow}c^\dagger_{\beta,\uparrow}c^{\phantom{\dagger}}_{\beta,\downarrow}c^{\phantom{\dagger}}_{\alpha,\uparrow}
+
c^\dagger_{\beta,\uparrow}c^\dagger_{\beta,\downarrow}c^{\phantom{\dagger}}_{\alpha,\uparrow}c^{\phantom{\dagger}}_{\alpha,\downarrow}
\! + \! h.c.) \,.
\end{multline} 
Here, $c^\dagger_{\alpha,\sigma}$ ($c_{\alpha,\sigma}$) creates
(annihilates) an electron with spin $\sigma$ in orbital $\alpha$;
$n_{\alpha,\sigma}=c^\dagger_{\alpha,\sigma}c^{\phantom{\dagger}}_{\alpha,\sigma}$.

In contrast for the $e_g$ orbitals, which are proportional
to $3z^2-r^2$ and $\sqrt{3}(x^2-y^2)$, the relation
 $U= U'+2J$ still holds for cubic symmetry: Again because of cubic symmetry
($x \leftrightarrow y$) any term involving one or three $x^2-y^2$ orbitals
vanishes; only
 the $U=U_{\alpha\alpha \alpha\alpha}$, $U'=U_{\alpha\beta \beta\alpha}$, 
 $J=U_{\alpha\beta \alpha\beta}=U_{\alpha \alpha \beta\beta}$ terms remain.
However now, instead of interchanging the orbitals,
 cubic symmetry operations  such as ($x \rightarrow x$, $y \rightarrow z$, 
$z \rightarrow -y$), lead to mixed orbitals:
$3z^2-r^2 \rightarrow -1/2 (3z^2-r^2) - \sqrt{3}/2  \sqrt{3}(x^2-y^2)$.
Hence, the intra-orbital Hubbard interaction $U$ for the
$e_g$ orbitals is not a cubic invariant, and $U$ has to depend on the other parameters $U'$ and $J$ through $U=U'+2J$.

\section{Quantitative deviations for {S\lowercase{r}VO$_3$} and
  {B\lowercase{a}O\lowercase{s}O$_3$}}
\label{Sec:numeric}

\subsection{Construction of Wannier functions}

We now aim to validate our analytical results and quantify the
deviation from the spherical-symmetry relation Eq.~(\ref{Eq:U}) in
real materials.  To this end, we perform DFT calculations\cite{Wien2k}
using a generalized-gradient approximation to the exchange-correlation
functional \cite{PBE} for two cubic perovskite materials, and
construct low-energy effective models using maximally-localized
Wannier functions (MLWF) \cite{Wannier90, wien2wannier}.

In terms of the formalism of Sec.~\ref{Sec:analytic}, the role of the
Wannier functions is to provide the radial dependence of the orbitals
which was irrelevant for our arguments from symmetry, but which must
be provided to compute numerical values for the interaction
parameters. The main difference is that we considered a local
octahedron before, while Wannier functions $|w_{\alpha\RR}\rangle$
properly belong to a periodic crystal: they have finite hopping
amplitudes $t_{\alpha\RR\alpha'\RR'}$ also for $\RR\neq\RR'$ (or
equivalently, they show a \kk-dispersion), and form an orthonormal set
$\langle w_{\alpha\RR}|w_{\alpha'\RR'}\rangle =
\delta_{\alpha\alpha'}\delta_{\RR\RR'}$ with respect to sites \RR and
orbitals $\alpha$.

Our first example is \svo, which is often used as a “testbed”
strongly-correlated material (for DFT+DMFT calculations, see
e.g. Ref.~\onlinecite{DMFTSrVO3 , DMFTSrVO3-2}; detailed discussions of Wannier
projections in this and related materials have been given in
Refs.~\onlinecite{pavarini,wien2wannier,scaramucci}).  The cubic perovskite  \svo is a
paramagnetic, correlated metal with electronic configuration $3d^1$,
i.e. one of the \ttg-derived states will be filled.

Secondly, we consider the recently synthesized compound
\boo. \cite{BaOsO3} With a low-spin $5d^4$ configuration, this is
another paramagnetic metal.  Since the $5d$ states are more extended
than the $3d$ states of \ce{V}, we expect to find greater $p$-$d$
hybridization and, in turn, greater deviation from $U=U'+2J$ in this
case.

For each material, we construct two sets of Wannier functions:
\begin{enumerate}
\item three ``$d$-only'' Wannier functions corresponding to the
  $\dd_{ij}$ of Eq.~(\ref{Eq:dprime}), and
\item twelve ``$d+p$''   Wannier functions corresponding   to the
  atomic $d_{ij}$ and $p_i$ states.
\end{enumerate}

It is instructive to compare these two approaches: The first set of
Wannier functions translates the three \ttg-derived bands to three
orbitals $|w'_{\alpha\mathbf 0}\rangle$ centered on the \ce{$B$} ion.
Direct and \ce{O}-mediated hopping processes are subsumed in an
effective $B$-$B$ hopping $t_{\dd\dd}$.  To account for this, the
$|w'\rangle$ must have substantial weight not only at the \ce{$B$} but
also at the \ce{O} atoms.  (In a band picture, the reason is the
significant \ce{O}-$p$ contribution to the \ttg-derived bands.)  Which
combinations of \ce{O}-$p$ and \ce{$B$}-$d$ orbitals mix is determined
by symmetry as discussed in Sec.~\ref{Sec:analytic}, cf.\
Fig.~\ref{Fig:DOS}.  Going beyond an effective single-particle
description, the Coulomb interaction is expected to be well
represented by a site local ``multi-band Hubbard'' term
$U_{\alpha'\beta'\beta\alpha}$ which can be parameterized by three
independent quantities $U$, $U'$, and $J$, as we saw in the previous
section.

The second set of Wannier functions spans nine additional bands, three
$p$-derived bands per \ce{O}.  With the $p$-states explicitly
included, the $d$-like MLWFs are free to become more localized; the
weight at the \ce{O} sites will be carried by the $p$-like orbitals
(cf.\ Fig~\ref{Fig:d+p}).  The downside is that the resulting model
becomes significantly more complex, since a correct treatment of such
a ``$d+p$'' model must take into account not only the intra-atomic
interactions on the \ce{$B$} ($U_{dd}$) and on the \ce{O} ($U_{pp}$)
sites, but also inter-atomic ($U_{pd}$) interactions. \cite{Nico,d+p}
This added complexity will increase the computational cost to solve
the model in any numerical method, but it will also make the physical
interpretation of the results more involved.

Before we turn to the results, note that the heavy ($Z=76$) element
\ce{Os} leads to an appreciable spin-orbit splitting in \boo.  Because it
would invalidate the symmetry analysis of Sec.~\ref{Sec:analytic}, we
neglect this effect in the construction of the Wannier functions.
While our analysis could be extended to include spin-orbit coupling, a
spin-orbit interaction term can also be added to the tight-binding
model afterwards in any case \cite{Zhong}.

\subsection{Results}

\begin{figure}
  \centering
  \raisebox{-0.59\height}{\includegraphics[width=.58\linewidth]{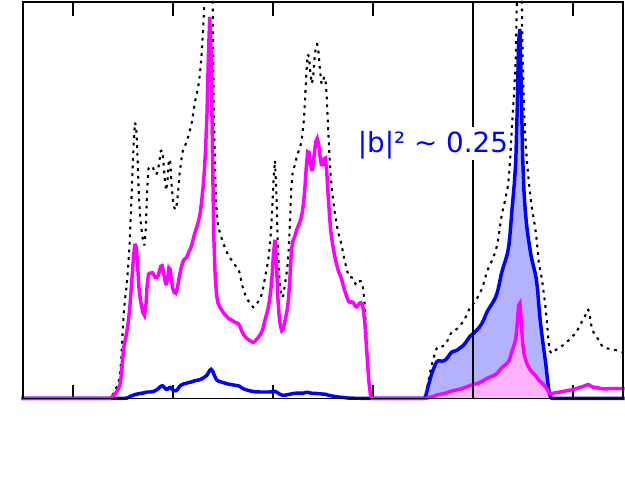}}
  \hfill
  \raisebox{-0.5\height}{\includegraphics[width=.38\linewidth]{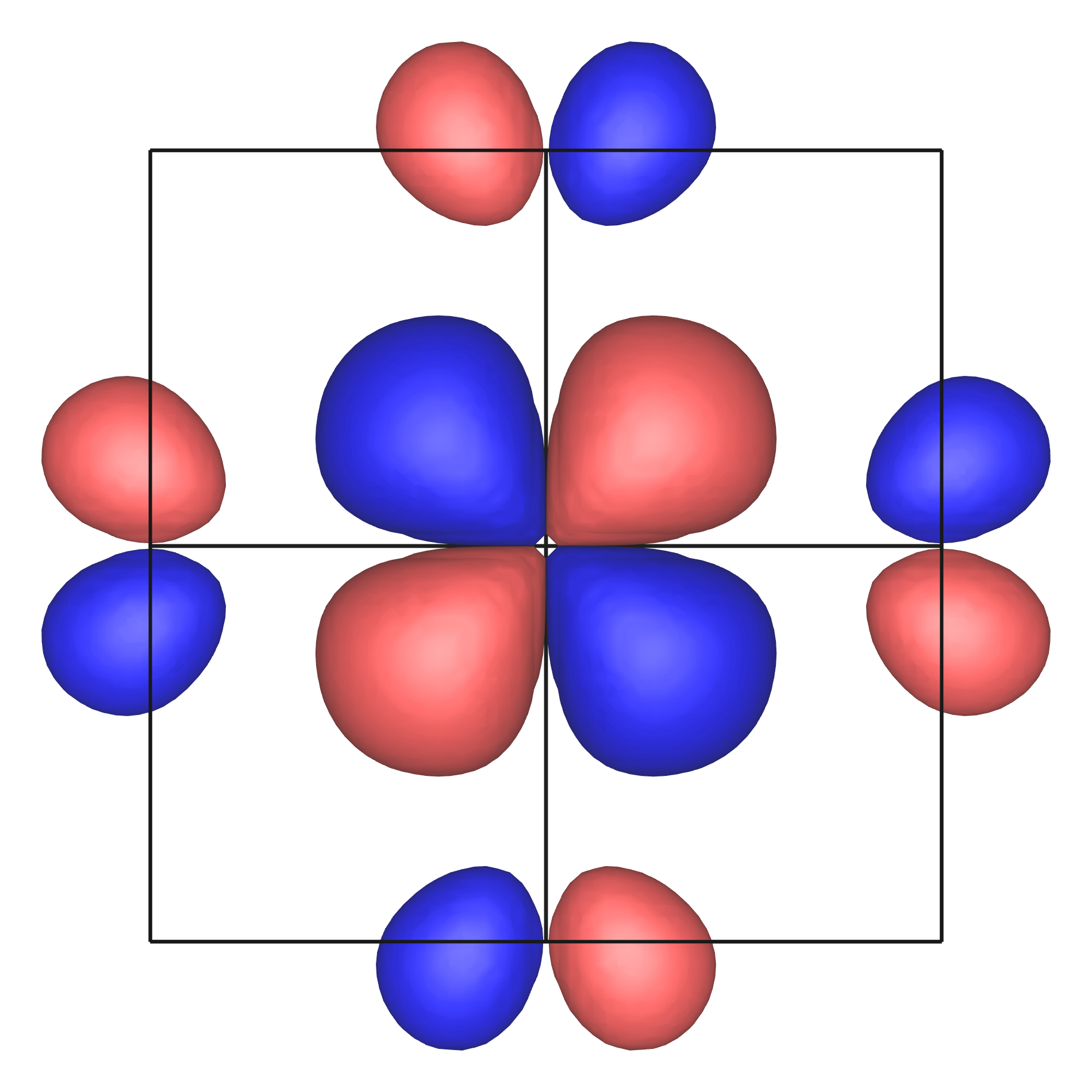}}
  \hspace*{\fill}
  \\[-5mm]
  \raisebox{-0.59\height}{\includegraphics[width=.58\linewidth]{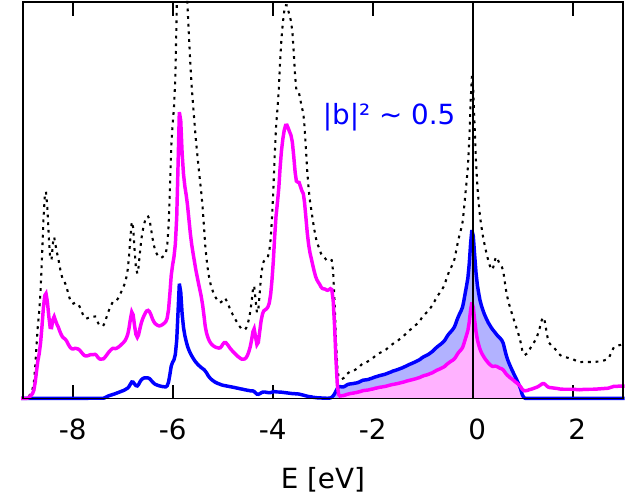}}
  \hfill
  \raisebox{-0.5\height}{\includegraphics[width=.38\linewidth]{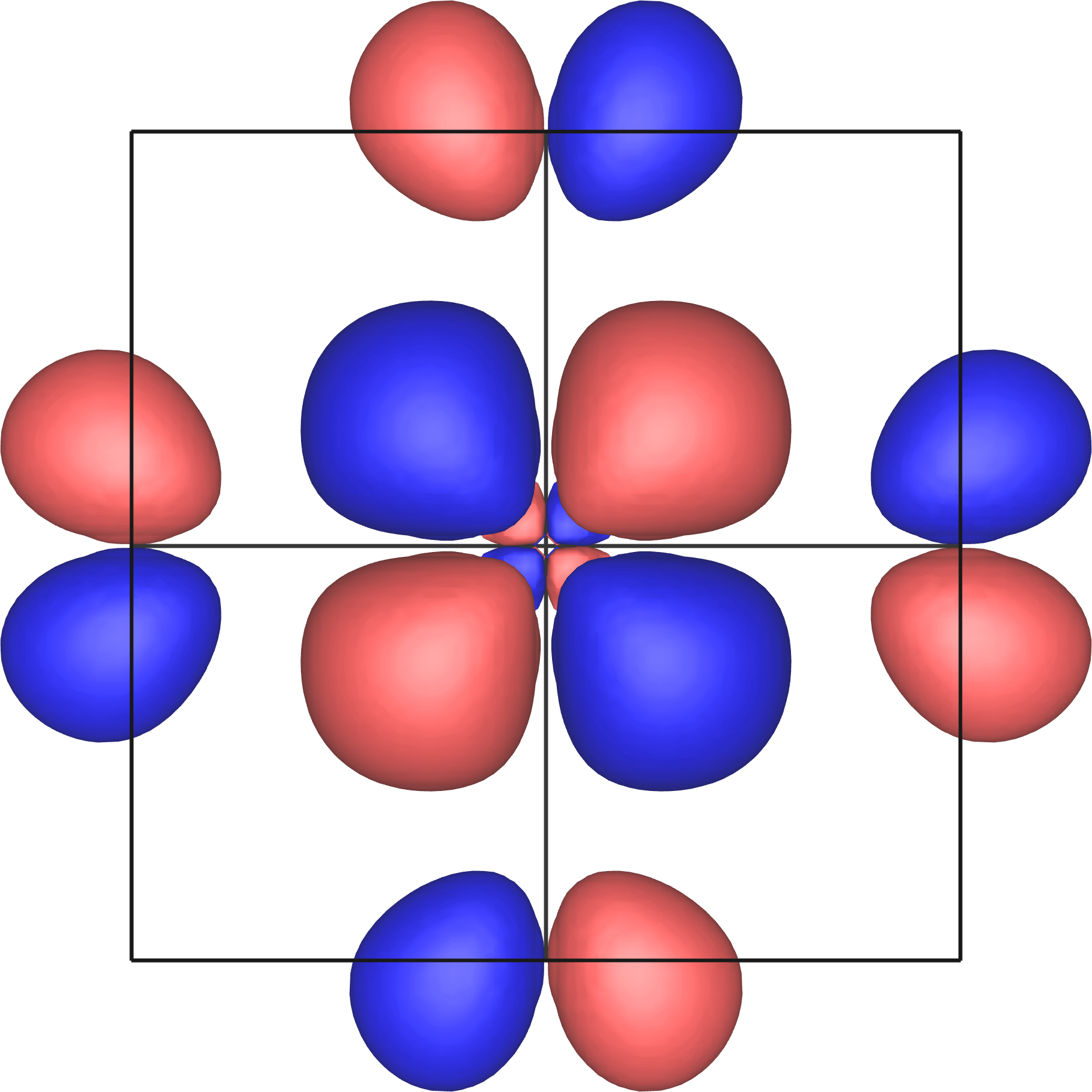}}
  \hspace*{\fill}
  \caption{(Color online.)  Densities of states and Wannier functions
    for the ``$d$-only'' Wannier orbitals for \svo (top) and \boo
    (bottom).  On the left, we show the total DOS (dotted line), and
    the projections on transition-metal \ttg (dark) and \ce{O} (light)
    states.  The shaded area marks the region of integration used to
    estimate the \ce{O}-$p$ weight [corresponding to $|b|^2$ in
    Eq.~(\ref{Eq:dprime})], see text.  On the right, the light / dark
    lobes are isosurfaces for the positive / negative parts of the
    real-valued Wannier functions.  The strong $p$-\ttg hybridization
    and the antibonding character are plainly visible.}
  \label{Fig:DOS}
\end{figure}

Fig.~\ref{Fig:DOS} shows the densities of states (DOS) of \svo and
\boo, and the Wannier orbitals corresponding to the 3-band case.  The
DOS around the Fermi level is derived from the $\pi$-antibonding
combinations of \ce{O}-$p$ and \ce{$B$}-\ttg states; correspondingly,
the 3-band orbitals are composed of a $d$-like contribution at the
\ce{$B$} site and $p$-like contributions at the \ce{O} sites sharing a
plane with the $d$-like part, akin to the $\oo_{\alpha}$ orbitals in
Section \ref{Sec:analytic}. These Wannier orbitals are also referred
to as ``$d$-only'', where the quotation marks hint that these orbitals
are actually not pure $d$-orbitals.  The Wannier functions are
equivalent to each other under cubic symmetry.

As expected, the $p$-$d$ hybridization is stronger in \boo than \svo.
This is seen both in the DOS (more \ce{O} weight around the Fermi
energy $E_F=0$) and in the orbitals (bigger lobes at the \ce{O}
sites).  We can quantify this observation by integrating over the
shaded areas in the DOS; this yields an \ce{O} admixture of
$|b_\svo|^2 \sim 0.25$ and $|b_\boo|^2 \sim 0.5$, respectively.  In
this sense, the ``$d$-bands'' of \boo consist in fact of almost equal
parts \ce{O} and \ce{Os} contributions.  These values agree
qualitatively with Eq.~\ref{Eq:dprime}, which yields $|b_\svo|^2 \sim
0.20$ and $|b_\boo|^2 \sim 0.33$ using the parameters from
Table~\ref{Tab:Hopping12}.\cite{mixnote} Quantitative differences have
to be expected because (i) Eq.~(\ref{Eq:dprime}) holds for an isolated
octahedron instead of the periodic crystal, (ii) there are further
hopping integrals that would have to be considered, and (iii) the
partial DOS of Fig.~\ref{Fig:DOS} are only projections within the
muffin tin spheres \cite{Wien2k}.

\begin{figure}
  \includegraphics[height=0.4\linewidth]{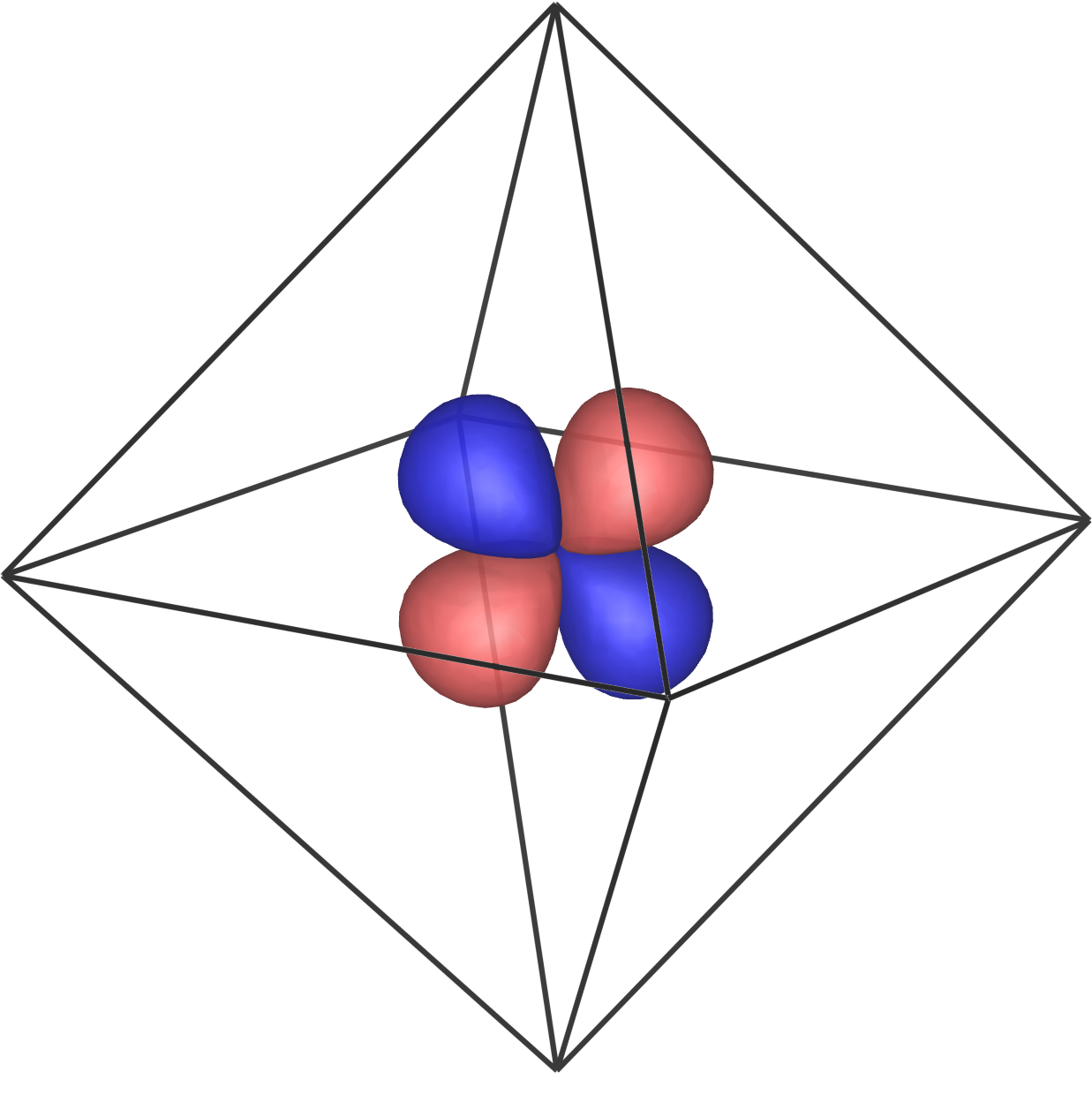}
  \hfill
  \includegraphics[height=0.4\linewidth]{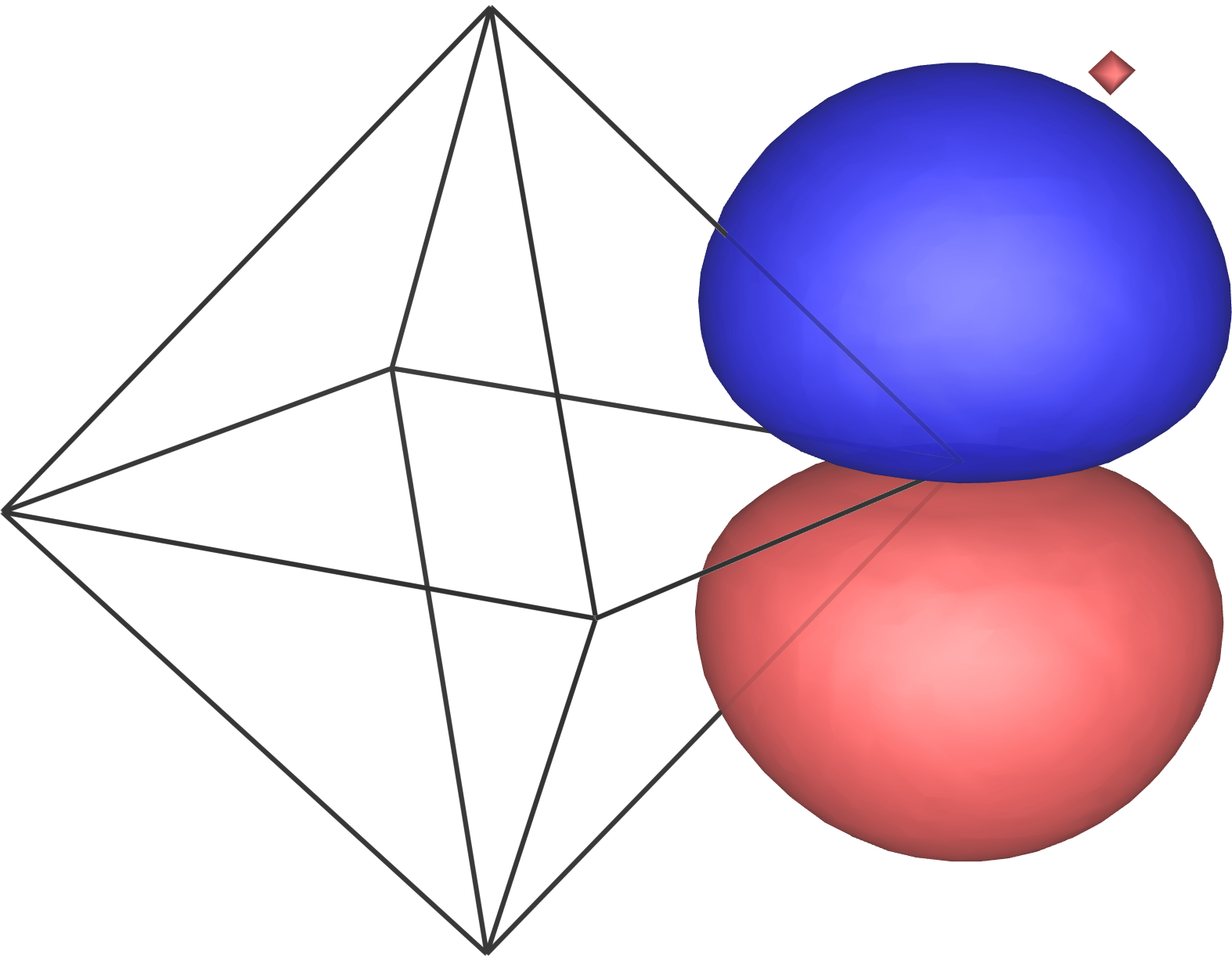}
  \\
  \hspace*{0.13\linewidth}
  \includegraphics[height=0.4\linewidth]{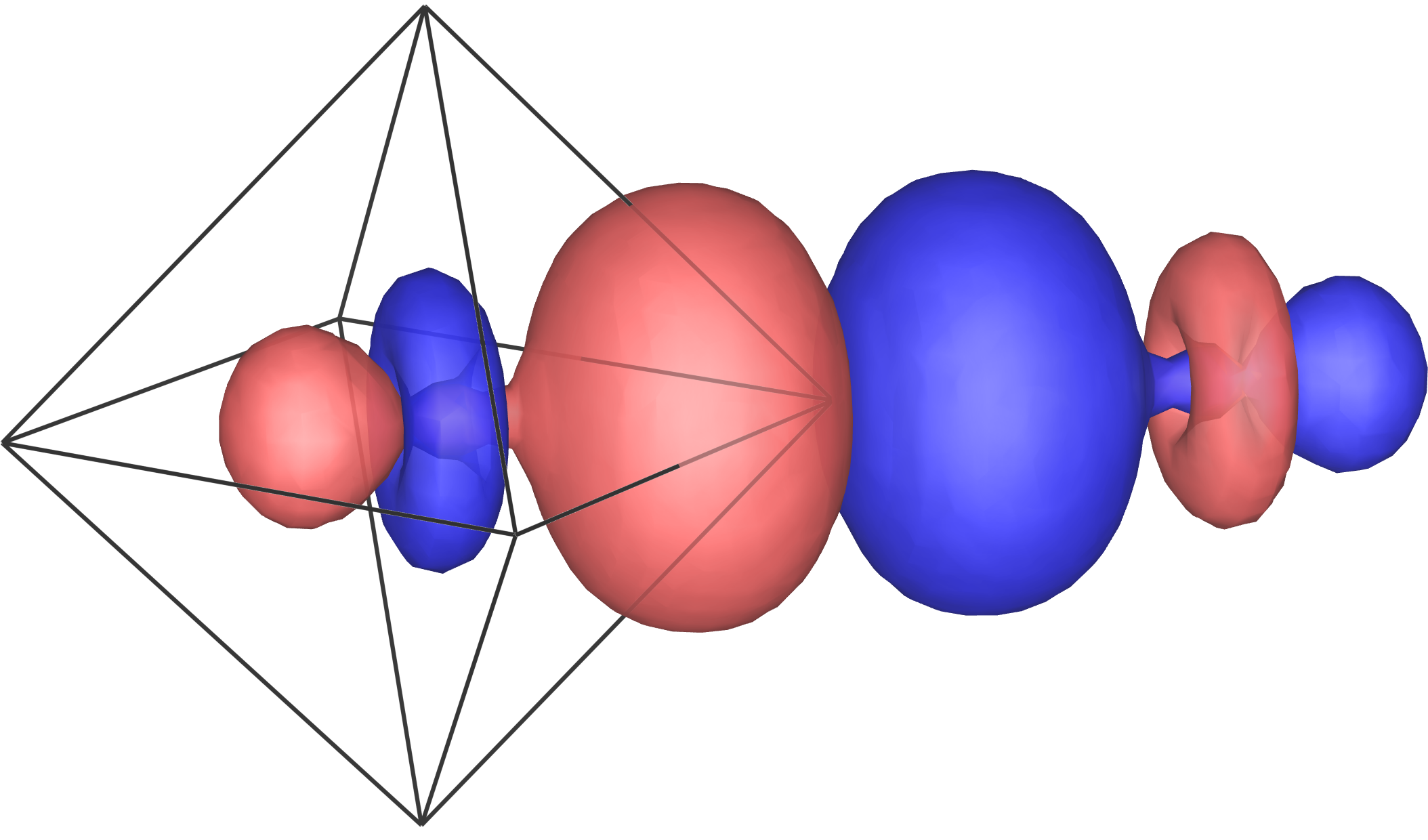}
  \caption{(Color online.)  12-orbital Wannier functions for \svo
    (plotted as in Fig.~\ref{Fig:DOS} right).  By symmetry, the twelve
    orbitals are grouped into three equivalent $d$-like orbitals (top
    left); and two types of $p$-like orbitals, three ``$p_\sigma$''
    (bottom) whose symmetry axes point toward their \ce{$B$}
    neighbors, and six ``$p_\pi$'' (top right) pointing away from the
    \ce{$B$} sites. With the \ce{O}-$p$ states explicitly included, no
    $p$-\ttg hybridization is seen in these orbitals.
    Correspondingly, the $d$ and $p_\pi$ orbitals are close to their
    atomic counterparts.  Conversely, the $p_\sigma$ orbitals, which
    mediate the $\sigma$-bonding between \ce{O}-$p$ and \ce{$B$}-\eg
    states, are elongated along their symmetry axis and have large
    contributions at their \ce{$B$}
    neighbors. \cite{pavarini,scaramucci}}
  \label{Fig:d+p}
\end{figure}

For these ``$d$-only'' Wannier orbitals, we have calculated the
Coulomb interaction by spatial integration of
Eq.~(\ref{Eq:U}). \cite{convnote} Table~\ref{tableU} summarizes the
results obtained for the bare interaction.  For the $3d^1$ perovskite
\svo, deviations from the spherical symmetric relation $U-U'=2J$ are
6\%. That is, calculations employing this relation can still be
expected to yield quite reliable results.  For the $5d^4$ perovskite
\boo, on the other hand, deviations are 25\%.  The reason for this is
the larger admixture of oxygen $p$ contributions, which according to
Section \ref{Sec:analytic} yields larger off-center Coulomb integral
overlaps and hence a larger deviation from spherical symmetry.

\begin{table}
  \caption{Coulomb interactions (unscreened) for ``$d$-only''
    Wannier functions of \svo and \boo; $(U-U')/2=J$ holds for
    spherical but not for cubic symmetry, for \boo deviations are
    indeed substantial.}
  \label{tableU}
  \smallskip
  \begin{tabularx}{\linewidth}{lX r@{.}l@{ eV}X r@{.}l@{ eV}}
    \hline\hline
    interaction &&
    \multicolumn{2}{c}{\svo} &&
    \multicolumn{2}{c}{\boo}
    \\ \hline
    $U$       &&  16&30 && 10&54\\
    $U'$      &&  15&14 &&  9&67\\
    $J$       &&   0&55 &&  0&33\\
    \hline
    $(U-U')/2$&&   0&58 &&  0&44\\
    \hline\hline
  \end{tabularx}
\end{table} 

Recently, transition metal oxides with heavy $4d$ or $5d$ elements
attract more and more attention. Indeed in such systems electronic
correlations are stronger than expected -- due to Hund's rule coupling
\cite{Georges13,Mravlje11,Hund}.  All the more important is a correct
Hamiltonian and multiplet structure with Hund's exchange. In this
respect, our finding highlights the substantial difference between
$(U-U')/2$ and $J$. A Kanamori Hamiltonian with three independent
Coulomb interactions needs to be considered for obtaining the correct
multiplet structure.

Next, we turn to the 12-orbital ``$d+p$'' Wannier functions.  This is
an alternative description of the low energy physics, where the oxygen
$p$-orbitals are explicitly taken into account. The corresponding
Wannier functions for \svo are shown in Fig.~\ref{Fig:d+p}.  The
$d$-like orbitals are again equivalent up to symmetry, but two
inequivalent types of $p$-like orbitals appear.  Symmetry also greatly
restricts the possible hopping processes between these states.  The
hopping amplitudes within the octahedron as well as selected
longer-ranged ones for all four Wannier projections are reported in
App.~\ref{App:hopping}.

We list the Coulomb interaction parameters between the 12-band
orbitals for \svo and \boo in Tables~\ref{tableUpdSVO} and
\ref{tableUpdBOO}, respectively.  For the $d$-like orbitals, $U=U'+2J$
is fulfilled with a reasonable accuracy of 3\% even in \boo.  Having
the additional degree of freedom regarding oxygen-$p$ Wannier
orbitals, the $t_{2g}$ orbitals are now localized around the
transition metal ion and have the spherically symmetric form,
cf.~Fig.~\ref{Fig:d+p}. In this case, two parameters are sufficient
for the $d$-$d$ Kanamori interaction.

As a side note, observe that in the 12-band case $U-U'<2J$, while in
the 3-band case $U-U'>2J$.  This is because $U'$ and $J$ are more
strongly reduced than $U$ by the shift of $t_{2g}$ weight to the
oxygens which occurs in the 3-band case, as $U'$ and $J$ are {\em
  inter-}orbital interactions that include more non-overlapping
oxygens in the interaction integral.

Let us emphasize that the $d$-$p$ interaction also plays an important
role \cite{d+p}.  The $d$-$p$ interactions of density-density type are
listed in Tables \ref{tableUpdSVO} and \ref{tableUpdBOO}
(right). There are two types of $p$ orbitals, denoted as $p_\pi$ and
$p_\sigma$ (see Fig.~\ref{Fig:d+p}). Interactions with $p$ orbitals
centered on an oxygen atom outside the plane of the $d$ orbital lobes
are denoted by $\perp $. The $d$-$p_\sigma$ interactions $U_{p_\sigma
  d}$ and $U_{p_\sigma d}^\perp$ with the $p_\sigma$ orbitals oriented
toward the transition metal site is considerably stronger than that
with the more regular $p_\pi$ orbitals. There is only one $U_{p_\pi
  d}^\perp$, while two parameters arise from density-density
interaction between $d$ and $p_\pi$ orbitals with the $p_\pi$ orbitals
being centered around oxygen sites within the plane defined by the $d$
orbitals. We denote these as $U_{p_\pi d}$ if the $p_\pi$ orbital lies
within the same plane and $U'_{p_\pi d}$ if it is oriented
perpendicular to it.  The considerable differences in the $d$-$p$
Coulomb interaction can be understood from the very different $p_\pi$
and $p_\sigma$ orbitals in Fig.\ \ref{Fig:d+p}. These differences are
of relevance for $d+p$ DFT+DMFT calculations that include
$U_{pd}$. \cite{d+p}

\begin{table}
  \caption{Left: Same as Table \ref{tableU} but for $d+p$  \svo
    Wannier functions. Right: the  different $d$-$p$ density-density
    Coulomb interactions for these Wannier functions, see main text
    for the notation.}
  \label{tableUpdSVO}
  \smallskip
  \begin{tabularx}{\linewidth}{lX r@{.}l@{ eV} XX lX r@{.}l@{ eV}}
    \hline\hline
    interaction && \multicolumn{2}{c}{\svo}
    &&&
    interaction && \multicolumn{2}{c}{\svo}
    \\\cline{1-4}\cline{7-10}
    $U$       &&  19&99 &&& $U_{p_\pi    d}$ &&		7&24\\
    $U'$      &&  18&52 &&& $U'_{p_\pi    d}$     &&  \phantom{1}7&18\\
    $J$       &&   0&74 &&& $U_{p_\sigma d}$     &&             8&52\\
    \cline{1-4}
    \multicolumn{4}{c}{ }&&& $U_{p_\sigma d}^\perp$ &&	6&87\\
    $(U-U')/2$&&   0&74  &&& $U_{p_\pi d}^\perp$	  && 	8&06\\
    \hline\hline
  \end{tabularx}
\end{table} 

\begin{table}
  \caption{Same as Table \ref{tableUpdSVO} but for $d+p$ \boo
    Wannier functions.}
  \label{tableUpdBOO}
  \smallskip
  \begin{tabularx}{\linewidth}{lX r@{.}l@{ eV} XX lX r@{.}l@{ eV}}
    \hline\hline
    interaction && \multicolumn{2}{c}{\boo}
    &&&
    interaction && \multicolumn{2}{c}{\boo}
    \\\cline{1-4}\cline{7-10}
    $U$       &&  14&90 &&& $U_{p_\pi    d}$ &&		6&94\\
    $U'$      &&  13&65 &&& $U'_{p_\pi    d}$     &&  \phantom{1}6&80\\
    $J$       &&   0&64 &&& $U_{p_\sigma d}$     &&             7&85\\
    \cline{1-4}
    \multicolumn{4}{c}{ }&&& $U_{p_\sigma d}^\perp$ &&	7&23\\
    $(U-U')/2$&&   0&62  &&& $U_{p_\pi d}^\perp$	   && 	6&38\\
    \hline\hline
  \end{tabularx}
\end{table}

\subsection{Effect of screening}
\label{Sec:screening}
The values reported above were calculated for a bare, unscreened
Coulomb interaction. In this section we include screening, within
Thomas-Fermi theory. That is, we replace the bare interaction in
Eq.~(\ref{Eq:U}) by
\begin{equation}
V(|{\mathbf r'}\!-\!{\mathbf r}|)=\frac{e^2}{4\pi \epsilon_0} \; \frac{1}{|{\mathbf r'}\!-\!{\mathbf
      r}|} e^{-|{\mathbf r'}\!-\!{\mathbf
      r}|/\lambda_{\rm TF}} \; ,
\end{equation}
where $\lambda_{\rm TF}$ is the Thomas-Fermi screening length. Let us
emphasize that for a cubic crystal the screened interaction
$V({\mathbf r},{\mathbf r'})$ itself will be of cubic symmetry, and
hence deviate from spherical symmetry. This effect is not taken into
account in the following; it will on its own generate further
deviations of the Kanamori interaction parameters from the spherical
relation $U=U'+2J$.

In the following, we adjust the parameter $\lambda_{\rm TF}$ to yield
a screened Coulomb interaction $U' \sim 3.5\,$eV for $3d$ \svo as
calculated by constrained LDA \cite{DMFTSrVO3 , DMFTSrVO3-2}.  This corresponds to a
screening length $\lambda_{\rm TF}=0.43$\AA{} (the lattice parameters
are $a_{\svo}=3.8425\,$\AA{} \cite{aSVO} and $a_{\boo}=4.025\,$\AA{}
\cite{BaOsO3}).  We employ the same screening length also for \boo as
this yields an interaction parameter $U'\sim 1.8\,$eV, which is in the
expected range for the $5d$ \boo.

Table \ref{tableUscreened} shows the results obtained for the
``$d$-only'' models.  In the case of $3d$ orbitals as exemplified by
\svo, deviations from spherically symmetric interaction parameters are
already small without screening and become negligible if screening is
included.  By contrast, for $5d$ \boo, $U -U' = 2J$ is significantly
violated even when screened.  Let us note that the degree of deviation
is quite robust over a large range of screening lengths.  For example
with $\lambda_{\rm TF}=0.61\,${\AA} we obtain a similar deviation of
14\% ($U = 2.55\,$eV, $U' = 1.90\,$eV, and $J = 0.28\,$eV).

\begin{table}
  \caption{Same as Table \ref{tableU} but for screened interaction
    with  screening length $\lambda_{\rm TF}=0.43$\AA.}
  \label{tableUscreened}
  \smallskip
  \begin{tabularx}{\linewidth}{lX r@{.}l@{ eV}X r@{.}l@{ eV}}
    \hline\hline
    interaction &&
    \multicolumn{2}{c}{\svo} &&
    \multicolumn{2}{c}{\boo}
    \\ \hline
    $U$       &&  4&40 && 2&44\\
    $U'$      && 3&47 &&  1&80\\
    $J$       &&   0&46 &&  0&28\\
    \hline
    $(U-U')/2$&&   0&47 &&  0&32\\
    \hline\hline
  \end{tabularx}
\end{table} 

Interestingly, for weak screening (large $\lambda$) $J$ can even be
enhanced whereas $U$ and $U'$ are always reduced.  The reason for this
is that the exchange integral $J$ includes positive {\em and} negative
contributions; and for large $\lambda_{\rm TF}$, the negative
contributions are more strongly reduced than the positive ones.  For
example at $\lambda_{\rm TF}=21.13$\AA{} we obtain $J=0.5465\,$eV for \svo,
which is larger than the unscreened $J=0.5464\,$eV. As the increase is very small, the results are given to a higher precision than elsewhere in the paper. With $U = 15.67\,$eV
and $U' = 12.50\,$ eV, deviations are 6.2\% for this screening strength.

In the limit of infinite screening, i.e., $\lambda_{\rm TF}\rightarrow
0$, one can show that $U' = J$.  That is, one can describe this limit
by one Kanamori interaction parameter $U=3U' = 3J$ for spherical
symmetry, and two ($U$ and $U'=J$) for cubic symmetry. Numerically, we
get however also for cubic ``$d$-only'' Wannier functions $U/J\sim 3$
for both \svo and \boo.  The limits of strong and weak screening show
that the idea that screening strongly reduces $U'$ and hardly reduces
$J$ is not true in general.  For strong screening, $J$ is reduced as
much as $U'$, since they are equal, while for weak screening $J$ is even enhanced.

\section{Conclusion}

We have analyzed the physical origin and the magnitude of the
difference between a spherically symmetric and a cubic interaction for
$t_{2g}$ orbitals. Deviations are quite large for $5d$ orbitals of
heavy transition metals. Since for these systems Hund's exchange is
paramount for electronic correlations,\cite{Georges13} we conclude that a Kanamori
interaction with three instead of two independent parameters is
necessary.  Unfortunately, this requires the calculation of one
additional interaction parameter and hence a more thorough analysis of
the interaction in DFT+DMFT calculations than customary hitherto.
Only if the oxygen degrees of freedom are included in the Wannier
projection, this is not necessary. In this case, however, the
(different) $d$-$p$ Coulomb interactions should be taken into account.
For $e_g$ orbitals there is no such difference between spherical and
cubic interaction.

Depending on the screening length, screening enhances or reduces the
difference between spherically symmetric and cubic interaction
parameters. Screening can even enhance $J$ whereas $U$ and $U'$ are
always reduced.  For Thomas-Fermi screening, $U = U'+2J$ is still
significantly violated for $5d$ \boo.  Let us note that the simple
Thomas-Fermi screening employed here is spherically symmetric, whereas
the physical screening function obeys the cubic, not the spherical
symmetry.  This effect is an additional source of deviations from
spherically symmetric interaction parameters.



For both $e_g$-only and $t_{2g}$-only low-energy effective models, we
have a Kanamori interaction for cubic symmetry. This
makes continuous-time quantum Monte-Carlo simulations \cite{ctQMC} very efficient because of
an additional local symmetry, see Ref.\ \onlinecite{PS}.

\acknowledgements{This work has been supported by the European
Research Council under the European Union's Seventh Framework
Programme (FP/2007-2013)/ERC through grant agreement n.\ 306447; and
by an {\em innovative project} grant from Vienna University of
Technology (EA).}

\appendix

\section{Coulomb interaction and Slater integrals}
\label{App:Slater}
For the sake of completeness, let us briefly add the representation of
the Coulomb interaction Eq.~(\ref{Eq:U}) by Slater integrals.
Expressing $1/|{\mathbf r} - {\mathbf r}'| = \displaystyle\sum_{l , m}
\dfrac{ {\rm min}(r,r')^l }{ {\rm max}(r,r')^{l+1} } \dfrac{4 \pi}{2l
  + 1} Y_{l, m}(\theta , \varphi) Y^*_{l , m}(\theta ' , \varphi ')$
in terms of spherical harmonics $Y_{l, m}$ and with
$\psi_\alpha({\mathbf r})=R(r) Y_{\alpha}$ where $R(r)$ is independent
of $l$ (or $\alpha$), the Coulomb interaction Eq.~(\ref{Eq:U})
becomes
\begin{widetext}
  \begin{multline}
    {U}_{\alpha' \beta' \beta \alpha} = \dfrac{e^{2}}{4 \pi \epsilon_0}
    \int \! {\rm d}r   {\rm d}\Omega  {\rm d}r'   {\rm d}\Omega'\,
    R(r) Y_{\alpha'}(\theta , \varphi) R( r ') Y_{\beta '}(\theta ', \varphi ')
    \\
    \cdot \displaystyle\sum_{l , m}  \left[  \dfrac{ {\rm min}(r,r')^l }{  {\rm
          max}(r,r')^{l+1} }
      \dfrac{4 \pi}{2l + 1}   Y_{l, m}(\theta , \varphi)  Y_{l , m}(\theta
      ' , \varphi  ') \right]
    R(r')  Y_{\beta }(\theta', \varphi')  R(r)  Y_{\alpha}(\theta ,
    \varphi)  {r'}^2 r^2 \; .
  \end{multline}
\end{widetext}
This integral can be decomposed into a radial ($ {\rm d}r {\rm d}r'$)and an angular part  (${\rm d}\Omega{\rm d}\Omega'$),
and the latter can be expressed in terms of  Clebsch-Gordan coefficients.
Thus only the radial integrals aka Slater parameters $F_l$ of
Eq.~(\ref{Eq:Slater}) need to be calculated.

\section{Hopping matrices}
\label{App:hopping}
In this Appendix, we report the numerical values of selected hopping
amplitudes in our Wannier projections.\cite{hopnote} The values for
\svo may be compared to Refs. \onlinecite{pavarini, wien2wannier}.
Note that this is not an enumeration of the largest hopping
amplitudes; rather, the selection is meant to be illustrative.

\begin{table}[tb]
  \centering
  \caption{Hopping amplitudes $t$ between the three $t_{2g}$ Wannier
    functions at various distances, and their on-site energies $E$
    relative to the Fermi level.  Values are in eV.}
  \smallskip
  \begin{tabularx}{\linewidth}{lX r@{.}lX r@{.}lX r@{.}lX r@{.}lX
      r@{.}lX r@{.}lX}
    \hline\hline
    && \multicolumn{2}{c}{$t^{(1)}_\pi$}
    && \multicolumn{2}{c}{$t^{(1)}_\delta$}
    && \multicolumn{2}{c}{$t^{(2)}_\sigma$}
    && \multicolumn{2}{c}{$t^{(2)}_\perp$}
    && \multicolumn{2}{c}{$t^{(2)}_\parallel$}
    && \multicolumn{2}{c}{$E$}
    \\ \hline
    \svo && $-$0&263 && $-$0&027 && $-$0&084 && 0&009 && 0&006 && 0&580
    \\
    \boo && $-$0&394 && $-$0&043 && $-$0&112 && 0&012 && 0&013 && $-$0&453
    \\
    \hline\hline
  \end{tabularx}
  \label{Tab:Hopping3}
\end{table}

For the 3-band Wannier functions (values in Table~\ref{Tab:Hopping3}),
no hopping is possible within the unit cell.  Two nearest-neighbor
hoppings are allowed, a $\pi$-type hopping $t^{(1)}_{\pi}$ when the
displacement is in the same plane as the orbital lobes (e.g. $xy$
orbitals with displacement $(1\, 0\, 0)$), and a smaller
$t^{(1)}_{\delta}$ of $\delta$ type where the displacement is
perpendicular (e.g. $xy$ and $(0\, 0\, 1)$).  Inter-orbital
nearest-neighbor hopping is forbidden by cubic symmetry.  There are
three second-nearest neighbor hopping parameters: $t^{(2)}_\sigma$
when both orbitals and the displacement share a plane (e.g. $xy$ and
$(1\, 1\, 0)$); $t^{(2)}_\parallel$ when the orbitals' planes are
parallel (e.g. $xz \leftrightarrow xz$ and $(1\, 1\, 0)$); and
$t^{(2)}_\perp$ when the planes are perpendicular (e.g. $xz
\leftrightarrow yz$ and $(1\, 1\, 0)$).

\begin{table*}[tb]
  \centering
  \caption{Selected hopping amplitudes $t$ between 12-band Wannier
    functions, and their on-site energies $E$ relative to the Fermi
    level.  Where the sign of the hopping alternates
    due to the signs of the $p$-type orbitals, we give the modulus.
    Values are in eV.}
  \smallskip
  \begin{tabularx}{\linewidth}{l r@{.}lX r@{.}lX r@{.}lX r@{.}lX
      r@{.}lX r@{.}lX r@{.}lX r@{.}lX r@{.}lX r@{.}lX r@{.}lX r@{.}lX
      r@{.}lX}
    \hline\hline
     & \multicolumn{2}{c}{$t_{dd\pi}^{(1)}$}
    && \multicolumn{2}{c}{$t_{dd\delta}^{(1)}$}
    && \multicolumn{2}{c}{$|t_{dp_\pi}|$}
    && \multicolumn{2}{c}{$|t_{p_\pi p_\pi}|$}
    && \multicolumn{2}{c}{$|t_{p_\pi p_\pi'}|$}
    && \multicolumn{2}{c}{$|t_{p_\pi p_\sigma}|$}
    && \multicolumn{2}{c}{$t_{p_\sigma p_\sigma}$}
    && \multicolumn{2}{c}{$t_{p_\pi p_\pi}^{(1)}$}
    && \multicolumn{2}{c}{$t_{p_\sigma p_\sigma}^{(1)}$}
    && \multicolumn{2}{c}{$E_d$}
    && \multicolumn{2}{c}{$E_{p_\pi}$}
    && \multicolumn{2}{c}{$E_{p_\sigma}$}
    \\
    \hline \svo
     & $-$0&128
    && $-$0&005
    &&   1&099
    &&    0&064
    &&    0&369
    &&    0&258
    && $-$0&044
    && $-$0&078
    &&    0&671
    && $-$0&407
    && $-$3&780
    && $-$5&520
    \\
    \boo
     & $-$0&187
    && $-$0&002
    &&    1&240
    &&    0&007
    &&    0&204
    &&    0&195
    && $-$0&024
    && $-$0&107
    &&    0&903
    && $-$2&063
    && $-$3&896
    && $-$6&887
    \\
    \hline\hline
  \end{tabularx}
  \label{Tab:Hopping12}
\end{table*}

For the 12-band case (Table~\ref{Tab:Hopping12}), we report the
nearest-neighbor $d \leftrightarrow d$ hopping parameters
$t_{dd}^{(1)}$ analogous to those of the 3-band case, but not those to
further neighbors.  In any case, \ce{O}-mediated hopping, which was
subsumed in the hoppings of the 3-band orbitals, now has to be taken
into account explicitly.

Within the octahedron, the following hopping processes are possible:
$t_{dp_\pi}$ when the $p_\pi$ orbital resides in the plane defined by
the $d$ (e.g. $d_{xy} \leftrightarrow p_y^{+x}$); $t_{p_\pi p_\pi}$
between nearest \ce{O} neighbors, i.e. along an edge of the octahedron
(e.g. $p_y^{+x} \leftrightarrow p_y^{+z}$); $t_{p_\pi p_\pi'}$ which
is the same as the last, but between orbitals of different orientation
(e.g. $p_y^{+x} \leftrightarrow p_x^{+y}$); $t_{p_\pi p_\sigma}$ along
an edge (e.g. $p_y^{+x} \leftrightarrow p_y^{+y}$); $t_{p_\sigma
  p_\sigma}$ along an edge (e.g. $p_x^{+x} \leftrightarrow p_y^{+y}$);
$t_{p_\pi p_\pi}^{(1)}$ across the octahedron (e.g. $p_y^{+x}
\leftrightarrow p_y^{-x}$); and $t_{p_\sigma p_\sigma}^{(1)}$ across the
octahedron (e.g. $p_x^{+x} \leftrightarrow p_x^{-x}$).

Comparing the sequences of values for the two materials, the same
trends are observed (with the exceptions of $t^{(2)}_\perp \gtrless
t^{(2)}_\parallel$ and $|t_{p_\pi p_\pi}| \gtrless |t_{p\sigma
  p\sigma}|$).  However, the values for the ``$d$-only'' orbitals, and for
$dd$ and $pd$ processes in the 12-band orbitals, are in general larger
for \boo than \svo, the larger lattice constant of \boo
notwithstanding ($4.03\,\text\AA$ versus $3.84\,\text\AA$ for \svo).
This is reflective of the greater $p$-$d$ hybridization and spatial
extent of the $5d$ states.

Contrariwise, the 12-band $pp$ hopping processes have larger amplitude
in \svo.  Our interpretation is that in this case, the larger spatial
distance prevails; indeed, the difference in Wannier function spread
$\langle r^2 \rangle$ between \svo and \boo is much more pronounced
for the $d$ than for the $p$ orbitals.  The exceptions to this rule,
$t^{(1)}_{p_\pi p_\pi}$ and $t^{(1)}_{p_\sigma p_\sigma}$ (hopping
across the octahedron), may be explained by the stronger $p$-$d$
hybridization in \boo.


\begin{thebibliography}{10}
\bibitem{notecRPA}
Note that in constrained random phase approximation \cite{cRPA}
one can in principle and sometimes does calculate the full Coulomb interaction matrix. However, usually 
the subsequent DMFT calculation only employ Slater or Kanamori parameters
if not a Kanamori parameterization is assumed from the very beginning.\cite{DMFTSrVO3-2 , cRPA2}
\bibitem{rotnote} Note that without the $e_g$ orbitals the Hamiltonian
  itself is actually not spherically symmetric. Rotation of e.g. the
  $d_{xy}$ orbital around the $z$-axis maps it onto a linear
  combination involving (missing) $e_g$ orbitals; rotating it around
  the $x$- or $y$-axis maps it onto a linear combination of the
  (included) $t_{2g}$. The Hamiltonian is invariant under the latter
  rotations if $U=U'+2J$.
\bibitem{mixnote} Note that $t$ in Eq.~(\ref{Eq:dprime}) is $|t| \sim
  2|t_{dp_\pi}|$ since the former denotes the hopping between $d$ and
  $O$ orbitals, and the latter between $d$ and $p$.
\bibitem{convnote} We have checked for convergence and errors with
  respect to the number of Wannierization $k$-points, the size of the
  real space cell, the number of real space support points, and shifts
  of these support points. The accuracy of integrals is $\lesssim
  0.2\%$.
\bibitem{hopnote} The signs of the hopping amplitudes reported here
  are as they would appear in a Hamiltonian matrix, i.e. without the
  extra sign of ``$-t$''.
\end{thebibliography}
\end{document}